\documentclass[fleqn,usenatbib]{mnras}

\usepackage{newtxtext,newtxmath}

\newcommand\logl{$\log (L/L_\odot)$}

\usepackage{soul}

\usepackage[T1]{fontenc}

\DeclareRobustCommand{\VAN}[3]{#2}
\let\VANthebibliography\thebibliography
\def\thebibliography{\DeclareRobustCommand{\VAN}[3]{##3}\VANthebibliography}

\usepackage{graphicx}	
\usepackage{amsmath}	
\usepackage{multirow} 

\title[NGC~300 ULX-1: A Surviving Star]{JWST spectroscopy of SN~2010da/NGC~300 ULX-1: a surviving star hidden by dust}

\author[E. R. Beasor et al.]{
Emma R. Beasor,$^{1}$\thanks{E-mail: E.R.Beasor@ljmu.ac.uk}
Ryan Lau,$^{2}$
Nathan Smith,$^{3}$
Dominic J. Walton,$^{4}$
Marianne Heida$^{5}$
and Ben Davies$^{6}$
\\
$^{1}$Astrophysics Research Institute, Liverpool John Moores University, IC2, 146 Brownlow Hill, Liverpool, L3 5RF, UK \\
$^{2}$IPAC, California Institute of Technology, CA, USA \\
$^{3}$Steward Observatory, University of Arizona, 933 North Cherry Avenue, Tucson, AZ 85721-0065, USA \\
$^{4}$Centre for Astrophysics Research, University of Hertfordshire, College Lane, Hatfield, AL10 9AB, UK \\
$^{5}$National Institute for Public Health and the Environment (RIVM), P.O. Box 1, 3720 BA, Bilthoven, the Netherlands\\
$^{6}$Independent researcher \\
}

\date{Accepted XXX. Received YYY; in original form ZZZ}

\pubyear{\the\year{}}

\begin{document}
\label{firstpage}
\pagerange{\pageref{firstpage}--\pageref{lastpage}}
\maketitle

\begin{abstract}

We present new James Webb Space Telescope ($JWST$) NIRSpec and MIRI integral-field spectroscopy of the remarkable system SN~2010da / NGC~300 ULX-1, the only known ultraluminous X-ray source powered by a neutron star with a supergiant donor. Our new data, taken in November 2024, reveal that the optical and near-infrared counterpart has dramatically faded since 2018 and no longer exhibits molecular absorption features characteristic of a red supergiant. Instead, the spectral energy distribution shows the donor has returned to its pre-outburst appearance, and is dominated by infrared continuum consistent with an optically thick warm ($\approx$900~K) dust shell. The bolometric luminosity indicates the presence of a surviving luminous source with \logl=4.11$\pm$0.02. Radiative transfer modelling of the mid-infrared spectral energy distribution (SED) reveals a broad emission feature peaking near $\sim$11\,$\mu$m, best reproduced by silicon carbide (SiC) dust grains, a composition typically associated with carbon-rich evolved stars. We rule out a failed supernova scenario, which would predict a large drop in luminosity and continued fading. Given the association of SiC dust with AGB stars, we suggest the donor star may be an AGB star that has survived and is now heavily enshrouded — having returned to a dust-obscured state following a transient post-outburst phase in which the system appeared as a red supergiant. We propose a revised evolutionary timeline in which the 2010 outburst initiated sustained super-Eddington accretion and temporarily altered the circumstellar environment. These observations provide rare insight into eruptive mass loss, dust formation, and binary interaction in a unique system.

\end{abstract}




\section{Introduction}

Stellar evolution underpins all areas of astrophysics. Massive stars in particular have major influence on many fundamental galactic/extra-galactic observables, since their explosive deaths enrich the universe with heavy elements and drive chemical evolution. However, modern surveys and advances in stellar modeling are highlighting significant gaps in our understanding. While canonical theory suggests stars between 8 – 30 M$_\odot$  should all die as Type II-P H-rich supernovae before leaving behind a neutron star remnant, the apparent lack of high mass progenitors \citep{smartt2015observational, davies2020red,davies2020on} and the higher-than-expected detection rate of of binary stellar mass black holes \citep[BHs,][]{abbott2017gw1} are leading many to speculate that enhanced mass-loss phases could alter the evolutionary path and peel away the envelope of a star, forcing it to die as a H-poor Type Ibc SNe \citep[e.g.][]{georgy2012yellow}. However, observational estimates of wind mass-loss rates are lower \citep{smith14,beasor20} than mass-loss prescriptions used in most stellar evolution models, and observations of massive stars in galaxies and clusters show that any enhanced mass-loss phases are far too rare and short-lived to impact RSG envelopes \citep{beasor2022dersgs}. To further complicate matters, stars in this mass range are highly likely to have evolved in a binary system \citep{sana2012binary,moe2017mind}, the ultimate outcomes of which are diverse.

Some objects in the universe do undergo massive eruptions through which large amounts of mass may be ejected, most notably ‘SN imposters’ and luminous blue variables (LBVs) {such as $\eta$ Carinae, P Cygni and SN 2009ip, among others (see \citealt{smith2026} for a recent review). While the cause of these eruptions remains debated, some of them (like $\eta$ Car) are likely to be massive star merger events} \citep{smith18}. These transients can reach peak magnitudes of M$_{V}$ $\simeq$ $-$10 to $-$15, but do not result in the destruction of the progenitor star \citep{smith2011observed}. While rare, these volatile phases may be able to significantly alter the evolutionary path of a star by removing ~10s of solar masses of material in a short space of time \citep{smith2006infra,smith2003mass}. As such, LBVs and SN impostors provide a unique opportunity to understand the influence of eruptive mass loss on stellar evolution. However, the physical mechanism that powers major LBV events remains unknown. 

The SN imposter SN 2010da has continued to behave in unexpected ways since its initial discovery in 2010. The event was detected in the nearby spiral galaxy NGC 300, at a distance of only 1.86 Mpc \citep[][]{dalcanton2009dist}. The visual-wavelength light curve and spectral evolution of the initial optical transient \citep{chornock2010} showed similarities to LBVs and other SN impostors \citep{smith2011lbv}. 
Shortly after the optical outburst, a compact, persistent X-ray source was identified at the same location, leading to speculation that the system may be a high-mass X-ray binary powered by accretion onto a compact object (HMXB, \citealt{binder2011,binder2016}). Further X-ray observations of the system revealed luminosity above 10$^{39}$ ergs/s placing the system above the threshold to be classified as an ultraluminous X-ray source (ULX; see \citealt{pinto2023ulx} and \citealt{middleton2023ulx} for recent reviews on ULXs). As such, the system was re-dubbed NGC 300 ULX-1. ULXs are the most extreme members of the X-ray binary population, and are ultimately powered by accretion from a stellar binary companion onto a compact object \citep[see][]{kaaret2017ulx}. Furthermore, in the case of ULX-1 coherent X-ray pulsations with a period of $\sim$30 s were subsequently discovered \citep{carpano2018}, unambiguously revealing that the accretor is a neutron star that was undergoing highly super-Eddington accretion ($L_{\rm X}$/ $L_{\rm edd}$ $\sim$ 20) during its ULX phase\footnote{The detection of coherent X-ray pulsations with a 30s period unambiguously identifies the compact object as an accreting neutron star. The extreme spin-up rate \citep{vasilopoulos2018} and the $\sim$0.25c outflow \citep{kosec2018} further confirm that the X-ray emission is powered by accretion onto a compact object rather than, for example, colliding wind shocks.}. This makes SN\,2010da/ULX-1 the only SN impostor event known to host a NS companion, and one of the few ULX pulsars known (e.g. \citealt{bachetti2014ulx}, \citealt{eksi2015}, \citealt{furst2016ulx}, \citealt{israel2017ulx}). The spin period of the NS showed an extreme rate of spin-up during this phase (\citealt{vasilopoulos2018}), as expected for accretion at extreme rates. Such extreme accretion is further supported by the $\sim$0.25$c$ outflow subsequently detected (\citealt{kosec2018}). Broadband X-ray data also revealed the potential presence of a cyclotron resonant scattering feature at $\sim$13\,keV, implying a magnetic field strength of $\sim10^{12}$\,G for the NS (\citealt{walton2018}), relatively typical of the field strengths seen in other NS HMXBs (see \citealt{caballero2012} for a review).

Other multiwavelength studies revealed additional optical, IR, and X-ray peculiarities compared to other SN impostors \citep{villar2016,lau2016}. These studies also revealed it to be dust enshrouded, and suggested that the source remaining after the 2010 outburst was 0.4 dex more luminous than the pre-2010 progenitor.  Notably, the progenitor was suggested as a yellow supergiant consistent with an initial mass around 10-12 $M_{\odot}$ \citep{villar2016}; such stars are considered to be too low in luminosity to undergo the LBV-like eruptions often invoked to explain SN impostors. Pre-outburst imaging at the location of ULX-1 reveal the progenitor was not an X-ray emitter, suggesting the 2010 outburst triggered a major shift in the system’s configuration, initiating super-Eddington accretion and the ULX phase (see \citet[][]{binder2020ulx1} for a review of the evolution of SN 2010da / NGC 300 ULX-1).

Spectral observations of the donor star taken in 2018 showed evidence for a cool atmosphere and expanding wind \cite[e.g. CO bandheads, TiO absorption features, see][]{heida2019discovery}. Fitting stellar atmosphere models to the spectrum showed the star resembled a red supergiant (RSG), with a temperature in the range 3650 $<$ Teff (K) $<$ 3900, while the K-band flux implies a luminosity of log(L/L$_\odot$) = 4.45 ± 0.10. This is distinctly different from the less luminous yellow supergiant progenitor.  Fitting the full X-shooter spectrum however required two additional components; a blue excess which can be attributed to a hot black body where Teff =  20,000K or a power law ($\alpha$ $\approx$ 4), likely due to reprocessed X-ray emission from the outer accretion disk, and a red excess that is likely the result of warm dust in the vicinity of the system \citep{heida2019discovery}. ULX-1 is the only X-ray pulsar system thought to contain an RSG, making the system unique. For one, there is clear potential that this system could be a progenitor for a compact object binary system (either NS-NS or NS-BH), depending on whether or not the RSG explodes as a SN or collapses directly to black hole (BH). The RSG-NS combination also places the system as a candidate Thorne-Żytkow Object (TŻO) progenitor, which is a ‘star within a star’ \citep[e.g.][]{thorne1975red,thorne1977stars,podsiadlowski1995evolution,hirai2022tzo, farmer2023tzo}, predicted to form when a NS is swallowed by an RSG or a red giant.

 Further deepening the mystery surrounding SN2010da / NGC300 ULX-1, the source has faded significantly across X-ray, optical and infrared wavelengths since 2018 \citep[see][and refs within]{chene2025ulx1}. \citet{chene2025ulx1} present X-ray and optical observations of SN~2010da/NGC~300~ULX--1, demonstrating that the system has exited its super-Eddington accretion phase. Continued monitoring with \textit{Swift}/XRT shows that the super-Eddington activity has ceased, while the most recent \textit{Chandra} detection in 2020 revealed a comparatively low X-ray luminosity of $L_X \approx 10^{37}\,\mathrm{erg\,s^{-1}}$, indicating that accretion onto the neutron star persists but is no longer supercritical. The optical detection obtained with \textit{Gemini} in 2024 shows that the source is more than 2.5 magnitudes fainter in the $i$-band relative to when the donor was spectroscopically identified as an RSG. Interpreting the concurrent decline in accretion rate and optical luminosity as a consequence of rapid donor-star evolution, \citet{chene2025ulx1} propose that the donor underwent a silent core collapse to form a BH, potentially resulting in the emergence of a nascent neutron star--black hole binary system.

 Here, we present new spectral data from SN2010da / NGC 300 ULX-1 taken with the ESO Very Large Telecope ($VLT$) and the James Webb Space Telescope ($JWST$). The paper is organised as follows. In Section \ref{sec:data} we discuss the observations and data from $VLT$  and $JWST$. In Section \ref{sec:results} we present the results of our analysis, including the luminosity of ULX-1 and a comparison to the progenitor system. In Section \ref{sec:discussion} we discuss the implication of these results and a suggest a possible evolutionary scenario. Finally, in Section \ref{sec:conclusions} we present our conclusions. 

\section{Observations and data}\label{sec:data}
\subsection{VLT}
NGC 300 ULX-1 was re-observed with the VLT X-Shooter instrument on 25 October 2023 as part of program ID 0109.D-0813 (PI: Heida). The data consist of a near-infrared spectrum covering 0.58–2.48~$\mu$m at a resolving power of $R \approx 11{,}400$, with a total exposure time of 3000~s.

\subsection{JWST}

\begin{figure*}
    \centering
    \includegraphics[width=\linewidth]{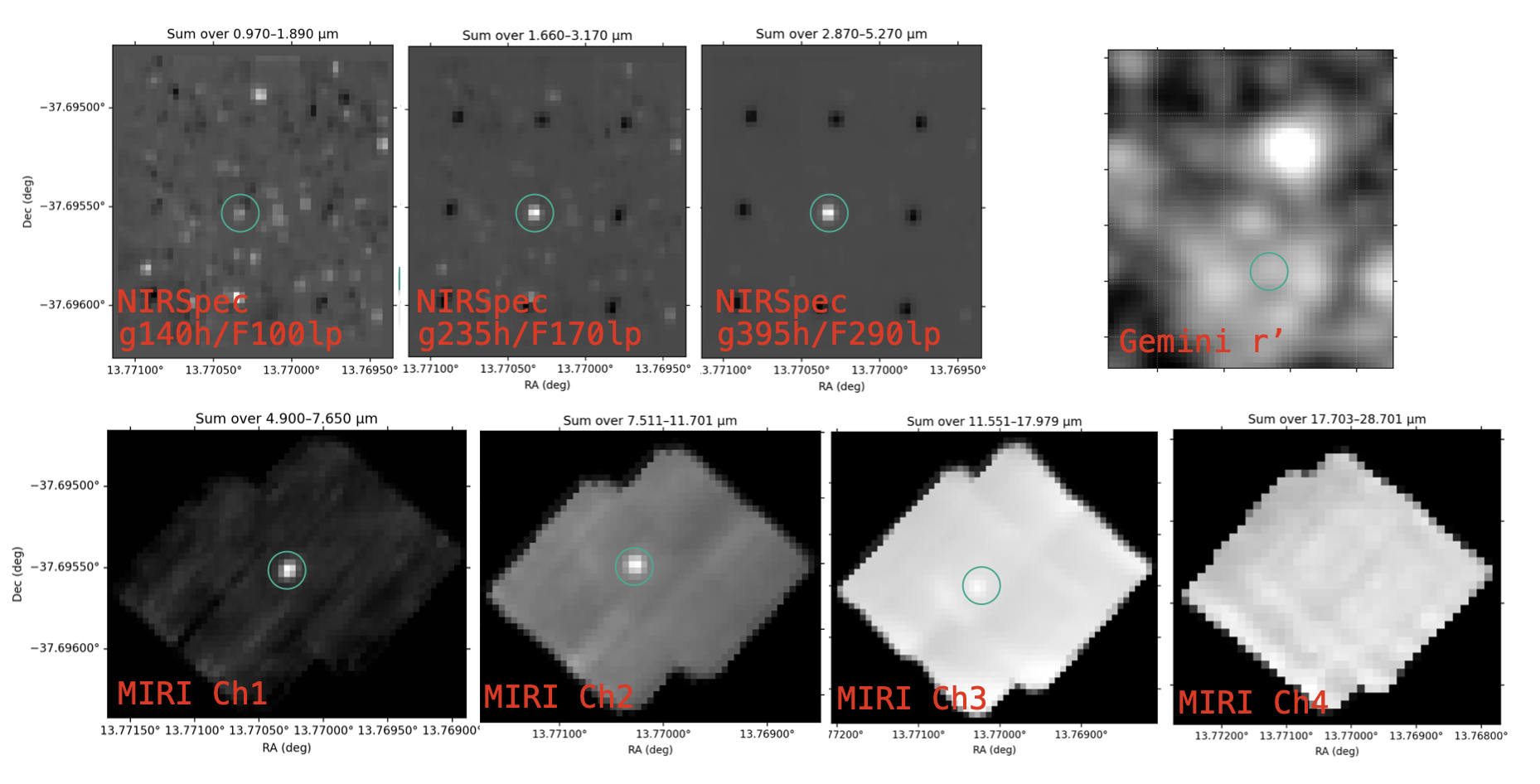}
    \caption{Collapsed IFU cubes for NIRSpec and MIRI. The location of the source is indicated with a green circle. In the top right we show a Gemini $r'$ band image taken in November 2024 \citep{chene2025ulx1}.}
    \label{fig:data}
\end{figure*}
Spectra of ULX-1 were taken as part of $JWST$ Cycle 2 program 3738 (PI: Beasor) on 11 November 2024. Observations were taken using the Mid-Infrared-Instrument (MIRI) in medium resolution spectroscopy (MRS) mode and the Near-Infrared Spectrograph (NIRSpec) in IFU mode. 

The NIRSpec IFU provides a $3^{\prime\prime} \times 3^{\prime\prime}$ field of view with spatial sampling of $0.1^{\prime\prime}$ per spaxel and enables simultaneous spatially resolved spectroscopy across the 0.6--5.3$\mu$m wavelength range. The grating/filters used were G140H/F100LP, G235H/F170LP, and G395H/F290LP, covering 0.97--5.27$\mu$m. The total on-source integration time was 5835.552~s. The target was centered in the IFU field using the standard NIRSpec target acquisition procedure. Observations were obtained using a four-point nod dither pattern to improve spatial sampling and mitigate detector artifacts.

The MIRI Medium Resolution Spectrometer (MRS) provides integral-field spectroscopy over the wavelength range 4.9--28.8~$\mu$m with a spectral resolving power of $R \sim 1500$--3500, depending on wavelength. The data were acquired in all four MRS channels, each comprising three sub-bands (A, B, C) to cover the full spectral range. The total on-source integration time was 2220.032~s. The target was centered in the MRS field of view using the dedicated target acquisition procedure. As for the NIRSpec observations, a four-point nod dither pattern was used.

 The Level 3 data were retrieved from the Mikulski Archive for Space Telescopes (MAST\footnote{https://mast.stsci.edu/}) and were reduced using the Calibration Reference Data System (CRDS) version 11.17.25. In both cases the spectra were extracted using {\tt jdaviz} software. The source spectra were extracted using circular apertures encapsulating the visible flux summed across all wavelengths. In Figure \ref{fig:data} we show the collapsed data cubes for each instrument and filter combination, as well as the location of the source compared to a recent Gemini $r'$ image \citep{chene2025ulx1}. 

\section{Results}\label{sec:results}

\begin{figure*}
    \centering
    \includegraphics[width=\linewidth]{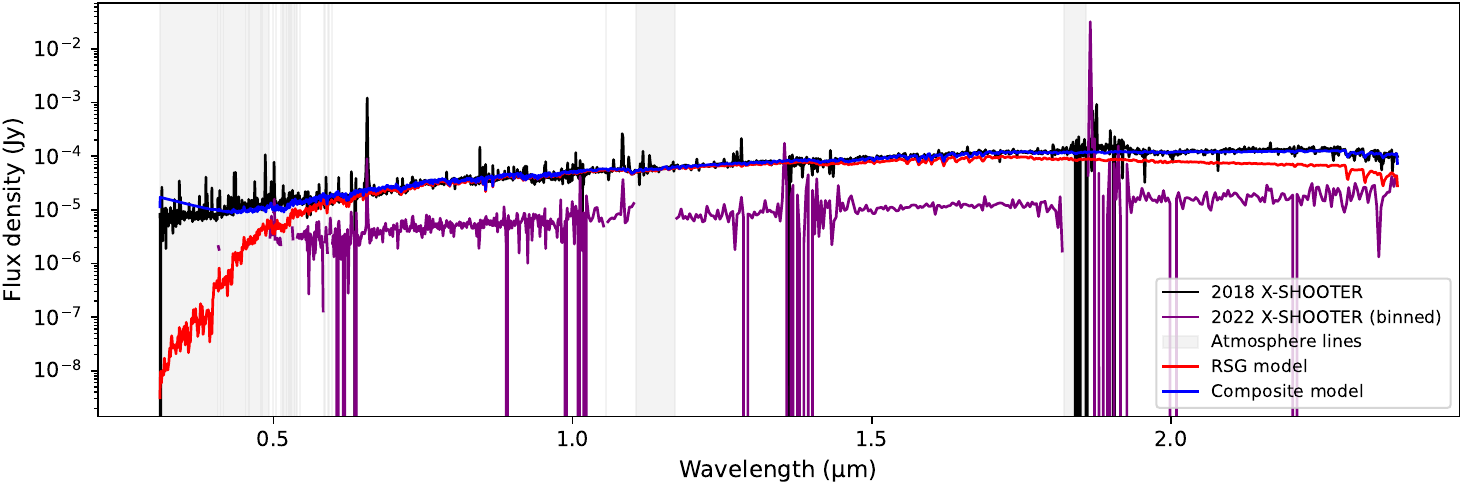}
    \caption{X-Shooter spectra of the donor star taken in 2018 (black line) and in 2022 (purple line), along with the best fit model to the 2018 spectrum presented in \citet{heida2019discovery} (red and blue lines). We mark the position of atmospheric absorption lines with grey bars. The 2022 spectrum has been binned to reduce noise.}
    \label{fig:xshooter}
\end{figure*}
\begin{figure*}
    \centering
    \includegraphics[width=\linewidth]{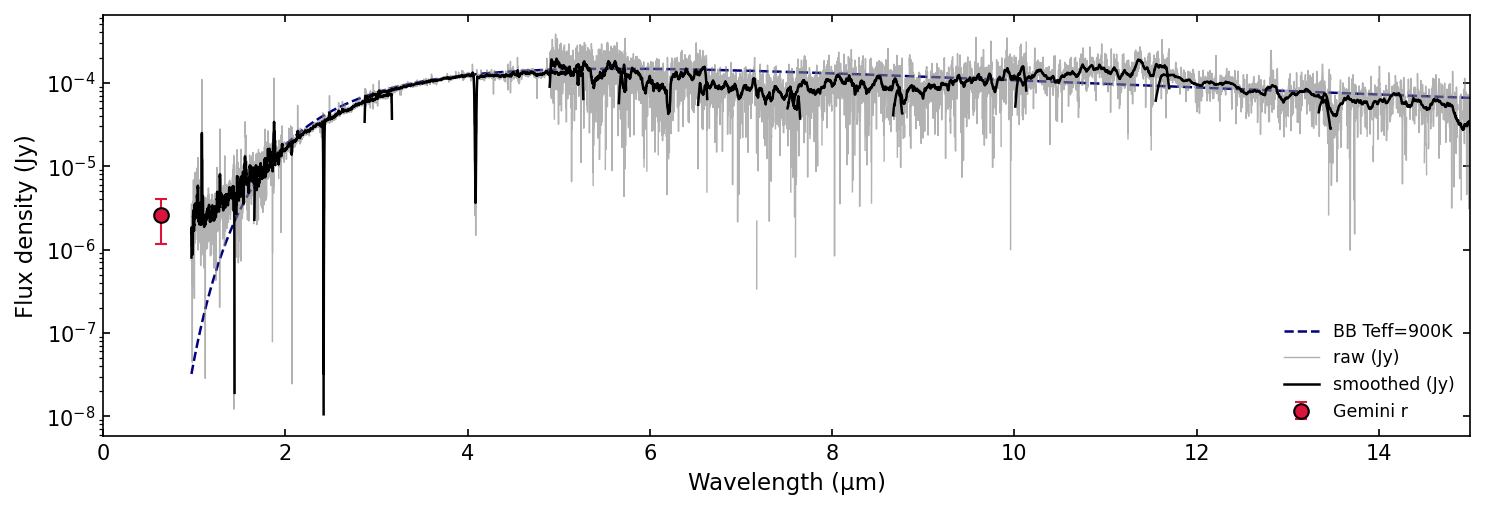}
    \caption{NIRSpec and MIRI spectra of ULX-1. We show the raw data (grey line) and smooth data (black line). The data have been smoothed using a moving average with a 50-pixel window for MIRI data and a 20-pizel window for NIRSpec data. We also overplot a 900K blackbody spectrum (dashed blue line) and the Gemini r' photometry point (red point) from \citet{chene2025ulx1}.}
    \label{fig:sed}
\end{figure*}

In Fig. \ref{fig:xshooter} we show the 2022 X-Shooter spectrum alongside the the 2018 spectrum for comparison. The 2022 spectrum is fainter than the 2018 spectrum by a factor of 10 between 0.6 and 2.5 $\mu$m. We no longer see the absorption features characteristic of a cool, extended atmosphere (TiO, CO) that were present in the 2018 spectrum. 

In Figure \ref{fig:sed} we show the extracted spectrum of ULX-1 including both $JWST$ NIRSpec and MIRI data. The newly observed SED appears significantly redder than the 2018 X-Shooter observations of the RSG-like counterpart to ULX-1.  Indeed, the flux at $\sim$1$\mu$m is fainter than the RSG-like spectrum by a factor of 10. Between 1 and 8$\mu$m, we find a cool, red component that can be approximately fit with a 900K blackbody spectrum\footnote{We note the absorption feature-like lines in the NIRSpec data are artifacts from the gap between CCDs.}. Beyond that, the SED appears to have a broad hump at 10-14 $\mu$m. We note however that this feature is narrower than a blackbody, and it also peaks at too long of a wavelength to be silicate emission (which peaks at around 10$\mu$m). 
n Fig \ref{fig:sed} we also show the Gemini $r'$ ($\lambda$ = 0.64$\mu$m) photometry presented in \citep{chene2025ulx1} with a red dot.

\subsection{Luminosity}
We now compare the luminosity of the observed infrared component to the luminosity of the RSG-like companion star reported by \citet{heida2019discovery}. To do this, we integrate under the observed SED and assume a distance of 1.86 Mpc \citep{dalcanton2009dist}, finding a luminosity of \logl= 4.10$\pm$0.01 . This is 56\% of the luminosity of the RSG-like source in 2018 \citep{heida2019discovery}, but is consistent with progenitor luminosity estimates from \citet{villar2016} and \citet{lau2016}, see Section~\ref{sec:progcomp}. We note that there is likely flux below our shortest wavelength point, and beyond our longest wavelength point. To compensate for this, we extrapolate the observed flux with a blackbody spectrum and include that in the luminosity calculation. For the blue flux, we extrapolate a 900K blackbody between 0 and 0.97$\mu$m. For the red flux, we extrapolate a 400K blackbody between 17.98 and 1000$\mu$m. We find that the missing flux contribution is 2.3\%, or 0.01 dex. To ensure the robustness of these assumptions, we also vary the two blackbody temperatures by $\pm$200K, finding this makes minimal difference on the derived luminosity. As such, our total luminosity is \logl=4.11$\pm$0.02. 

\subsection{Dust shell modeling}

We model the observed SED using the radiative transfer code {\tt DUSTY} \citep{ivezic1999dusty}. We assume the surviving central star is a cool supergiant \citep{villar2016} and explore a range of effective temperatures ($T_{\rm eff}$ = 2500, 3900, 5000, 7000~K). We created a grid of dust shell models varying the inner dust temperature ($T_{\rm in}$), optical depth ($\tau_{\rm V}$), grain size ($a$), and grain composition, as summarised in Table~\ref{tab:dusty_grid}.

The choice of grain composition strongly governs the shape of the mid-infrared SED. Pure silicate dust (Sil) produces a pronounced feature near $\sim$10\,$\mu$m, while pure graphite (grf) and amorphous carbon (amC) yield smooth, featureless continua that fail to reproduce the broad emission evident in the observed SED at $\sim$11\,$\mu$m. Mixed silicate and graphite compositions \citep{draine1984optical} suppress the silicate feature but do not produce emission at these wavelengths. Alumina (Al$_2$O$_3$) models similarly fail to reproduce the observed feature shape. 

\begin{figure}
    \centering
    \includegraphics[width=\columnwidth]{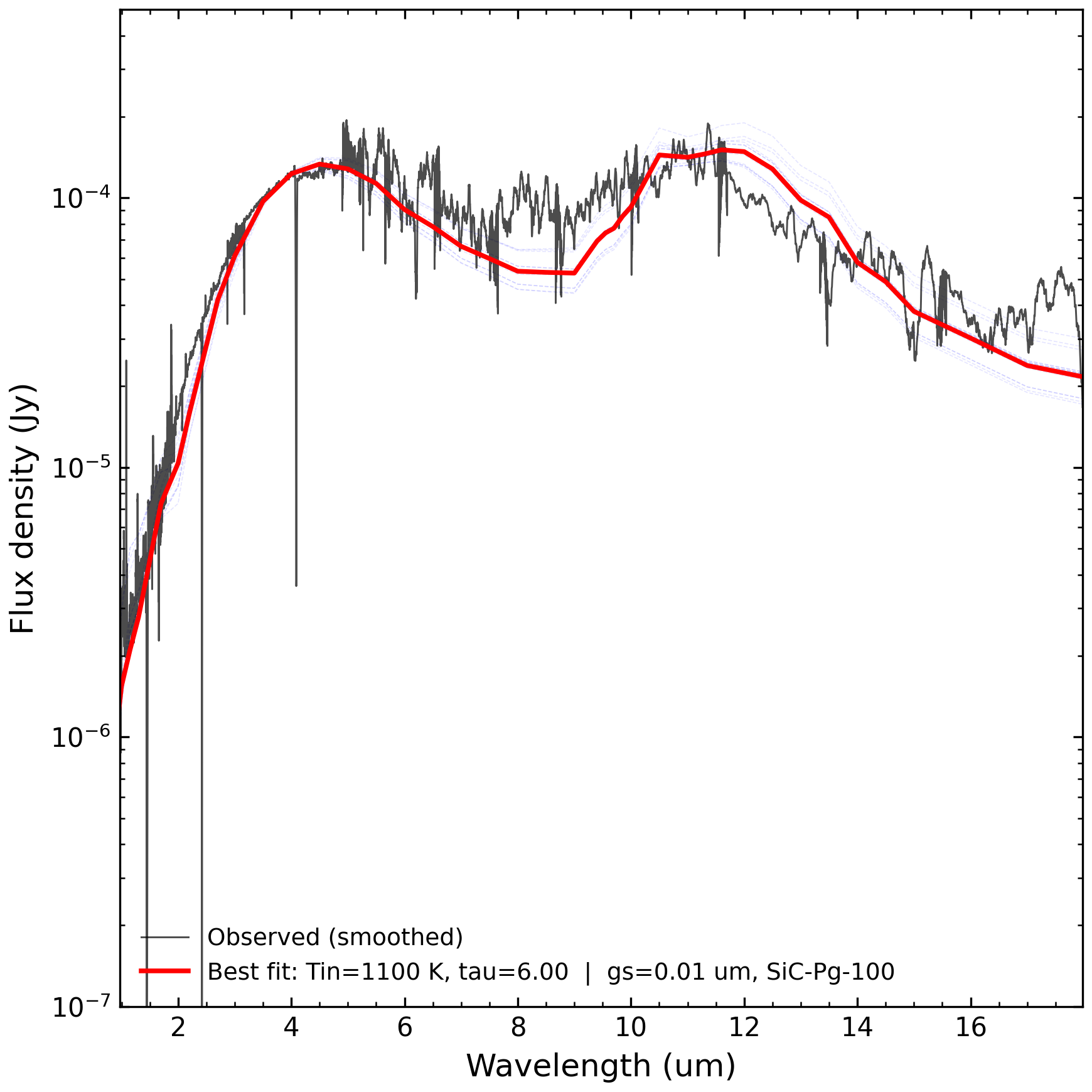}
    \caption{JWST NIRSpec and MIRI spectrum of SN\,2010da\,/\,NGC\,300 ULX-1 (black line)
with the best-fitting {\tt DUSTY} radiative transfer model overlaid (red line). The best-fitting model adopts SiC-Pg (silicon carbide) grains with $T_{\rm eff} = 3400$\,K, $T_{\rm in} = 1100$\,K, $\tau_{\rm V} = 6.0$, and $a = 0.01\,\mu$m. The dashed blue lines show all models within $\chi^2_{\rm min} + 3\sigma$, illustrating the range of dust shell parameters consistent with the observed SED. The shape of the broad emission feature peaking near $\sim$11\,$\mu$m is naturally reproduced by the SiC models, and is not reproduced by any other composition in our grid. The data have been smoothed using a moving average with a 50-pixel window for MIRI and a 20-pixel window for NIRSpec.}
    \label{fig:bestfit}
\end{figure}

In contrast, models incorporating silicon carbide grains (SiC) naturally reproduce the shape of the broad emission feature peaking near $\sim$11\,$\mu$m that is characteristic of the observed SED. In Fig.~\ref{fig:bestfit} we show our best-fitting {\tt DUSTY} model, which adopts SiC grains with $T_{\rm eff}$ = 3400~K, $T_{\rm in}$ = 1100~K, $\tau_{\rm V}$ = 6.0. The optical properties of the dust are described in \citet{pegourie1988}. We find that the best-fitting grain size is $a = 0.01\,\mu$m, smaller than the grains typically assumed for evolved star dust shells \citep[typically between 0.3 - 3$\mu$m,][]{amari1994}. Interestingly, the presence of SiC emission is typically associated with carbon-rich evolved stars, such as asymptotic giant branch stars  \citep[AGBs, e.g.][]{pegourie1988, sloan2024, boyer2025}. SiC grains are also among the most abundant presolar grains identified in meteorites, suggesting that carbon-rich stellar outflows may be a significant source of dust in the early Solar System \citep{Zinner2014, Nittler2016}.








\begin{table}
    \centering
    \begin{tabular}{l|c}
        Parameter & Range \\
        \hline
        $T_{\rm eff}$                        & 2500\,K, 3400\,K3900\,K, 5000\,K, 7000\,K \\
        $T_{\rm in}$                          & 400\,K\,--\,1200\,K, steps of 100\,K \\
        $T_{\rm in}$  (SiC-Pg only)           & 400\,K\,--\,2000\,K, steps of 100\,K \\
        $\tau_{\rm V}$                        & 1.0\,--\,15.0 (steps of 1.0) \\
        $a$                                   & 0.01\,$\mu$m, 0.3\,$\mu$m, 0.5\,$\mu$m, 1.0\,$\mu$m \\
        $a$ (SiC-Pg only)                     & 10$^{-4}$, 10$^{-3}$, 10$^{-2}$, 0.05, 0.1, 0.3\,$\mu$m \\
        \multirow{4}{*}{Composition}          & Sil-DL\,:\,grf-DL\,=\,100:0, 50:50 \\
                                              & amC-Hn\,=\,100 \\
                                              & SiC-Pg\,=\,100 \\
                                              & Al$_2$O$_3$\,=\,100 \\
    \end{tabular}
    \caption{Parameters of the {\tt DUSTY} model grid. All models assume a
    spherically symmetric dust shell with a density profile $\rho \propto r^{-2}$ and an outer radius $Y = r_{\rm out}/r_{\rm in} = 10^{4}$. Optical depths $\tau_{\rm V}$ are evaluated at $\lambda = 0.55\,\mu$m. For SiC-Pg models, an extended grain size grid was also explored. Sil-DL and grf-DL are the silicate and graphite grains of \citet{draine1984optical}; amC-Hn is amorphous carbon \citep[][]{hanner1988}, SiC-Pg is silicon carbide \citep[][]{pegourie1988}, and Al$_2$O$_3$ is alumina.}
    \label{tab:dusty_grid}
\end{table}
 
\subsection{Comparison to progenitor system}\label{sec:progcomp}
\begin{figure}
    \centering
    \includegraphics[width=\linewidth]{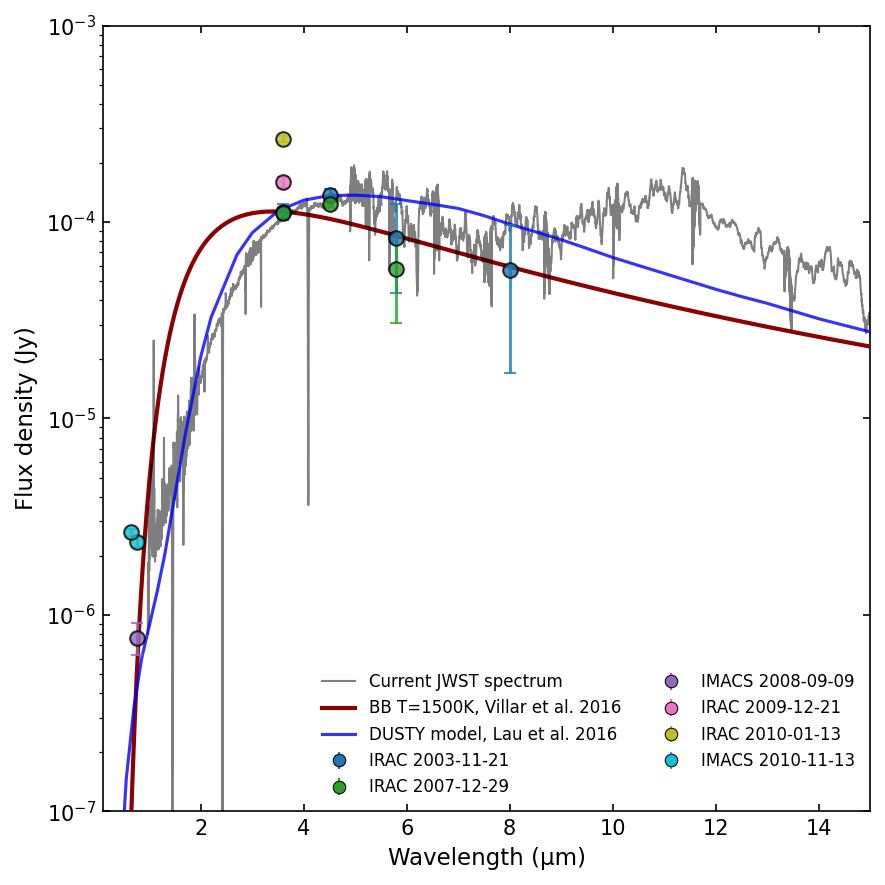}
    \caption{JWST spectrum with pre-outburst photometry overplotted. We also show the best-fit blackbody ($T_{BB}$ = 1500K) from \citet{villar2016} (red line) and the best fit {\tt DUSTY} model from \citet{lau2016} ($T_{\rm eff}$ = 7000K, $T_{\rm in}$ = 1000K, $\tau_{\rm V}$ = 5, blue line).}
    \label{fig:prog_comp}
\end{figure}

In Figure \ref{fig:prog_comp} we plot the JWST spectra together with the progenitor photometry from \textit{Spitzer}/IRAC and MMT/IMACS. The progenitor luminosity has previously been estimated by both \citet{villar2016} and \citet{lau2016}, who found bolometric luminosities of \logl=4.28$\pm$0.06 and \logl=4.18$\pm$0.06, respectively. We note that \citeauthor{lau2016} adopted a distance of 2 Mpc to NGC 300. When updating this to 1.86 Mpc, their inferred luminosity is \logl=4.11$\pm$0.06, consistent with the value reported in this work.  In \citet{villar2016}, the luminosity was derived from a blackbody fit to the progenitor SED, yielding a temperature of $\sim$1500 K. \citet{lau2016} modeled the progenitor photometry using radiative transfer calculations with \textsc{DUSTY} \citep{ivezic1999dusty}, finding that the SED could be fit by a star embedded in a dusty circumstellar shell ($T_{in}$=1000K, $\tau_{\rm V}$ = 5).

Importantly, in the progenitor SED a wide region around the peak of the blackbody was unconstrained because there was no photometry between $<$1 $\mu$m and 3.6 $\mu$m. With the continuous spectral coverage now available between 0.8 and 3.6 $\mu$m, we instead find the emission can be fit by a cooler temperature blackbody of $\sim$900 K (see Fig. \ref{fig:sed}). We note that the \textsc{DUSTY} model of \citet{lau2016} closely matches the appearance of the $JWST$ SED (see Fig. \ref{fig:prog_comp}). Overall, the close match between the JWST spectrum and the \citet{lau2016} model suggests that the luminous supergiant star has likely returned to its pre-outburst state, enshrouded by a cool dust shell.


\begin{figure*}
    \centering
    \includegraphics[width=475pt]{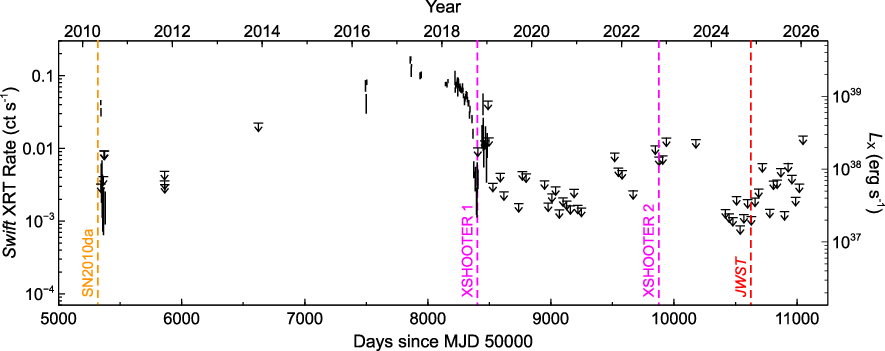}
    \caption{The X-ray lightcurve (0.3--10.0\,keV) of ULX-1 from the \textit{Neil Gehrels Swift Observatory} (\textit{Swift}; \citealt{SWIFT}) following its 2010 outburst. The data have been extracted using the standard online XRT pipeline (\citealt{Evans09}). During the main period of activity (prior to MJD 58500) the data are shown on a per-observation basis, while after this date we extract the XRT data using $\sim$month-long bins to better showcase the limits on continued activity placed by \textit{Swift}; where only upper limits on the brightness can be placed, these are computed at the 2$\sigma$ level. In addition to showing the evolution of the X-ray luminosity, we also indicate the date of the SN2010da event (that we take to mark the onset of the outburst) and the dates on which optical/IR spectra were taken with vertical dashed lines.}
    \label{fig:xraylc}
\end{figure*}

\section{Discussion}\label{sec:discussion}

We present new \textit{JWST} NIRSpec and MIRI IFU observations of the peculiar system SN~2010da / NGC~300~ULX-1. These data reveal that the source has returned to a heavily reddened, infrared-dominated state, closely resembling its pre-2010 progenitor. The bolometric luminosity remains consistent with earlier estimates. The optical flux is strongly suppressed and the near- and mid-infrared emission dominates the spectral energy distribution. 

Importantly, the current spectrum no longer exhibits the molecular absorption features (e.g.\ TiO, CO) that characterised the donor during its post-outburst phase when it resembled an RSG. Instead, the emission is well described by SiC dust reprocessing the luminosity of a deeply embedded central source. Together, these observations suggest that the RSG-like configuration observed between 2010 and 2018 did not represent the progenitor's long-term evolutionary state, but rather a transient phase following the 2010 outburst. The system has now returned to a dust-enshrouded configuration similar to that observed prior to the outburst. 

Notably, the current SED is better characterised by the lower luminosity and carbon-rich dust chemistry more typical of an AGB star than an RSG, raising the question of whether the donor is in fact a C-rich AGB star whose RSG-like appearance was a temporary consequence of the outburst. In this picture, the enhanced luminosity during the post-outburst phase — driven by reprocessed NS accretion luminosity (see Section~\ref{sec:transientrsg}) — may have inflated the photosphere sufficiently to mimic the cooler effective temperature and molecular absorption features of an RSG.

\subsection{A transient RSG phase induced by binary interaction}\label{sec:transientrsg}

The pre-2010 progenitor of SN~2010da was already known to be highly obscured by circumstellar dust, with the majority of its luminosity emerging in the infrared \citep{villar2016}. The appearance of a relatively unobscured RSG spectrum in the years following the 2010 transient therefore implies that the outburst temporarily altered the circumstellar environment and stellar structure of the donor.

We propose that the 2010 event disrupted or cleared the inner dusty envelope through a combination of radiative heating, shocks, and dynamical interaction, reducing the optical depth along the line of sight. Simultaneously, energy deposition into the outer stellar layers—likely associated with strong binary interaction—led to inflation of the donor's envelope, resulting in a cool, RSG-like atmosphere that we see in the 2018 X-Shooter spectrum (see Fig. \ref{fig:xshooter}). During this phase, the donor appeared optically bright and more luminous than the progenitor, and exhibited molecular absorption features characteristic of an extended, cool atmosphere.

\subsection{Re-enshrouding and the return to a dusty state}

The new \textit{JWST} observations indicate that the system has since re-entered a dust-enshrouded phase, as it appeared before its 2010 eruption. The disappearance of molecular absorption features seen in the 2018 X-Shooter spectrum and the dominance of warm dust emission imply that circumstellar material has re-formed or re-condensed around the donor, once again obscuring the stellar photosphere.

For a temperature of $\sim$900~K and a luminosity of \logl\ = 4.1, the characteristic radius of the emitting dust is of order $10^{14}$--$10^{15}$~cm. This scale is consistent with dust formation in a dense stellar wind or in recently ejected material that has cooled and become optically thick. The timescale of several years between the optically bright RSG phase and the current obscured state could be explained by dust condensation and wind-driven mass loss following the relaxation of the inflated stellar envelope. We therefore interpret the present appearance of ULX-1 as a return to the progenitor's original embedded state, rather than a new or terminal evolutionary phase.

We note that if the binary interaction responsible for the 2010 outburst is cyclical in nature, the pre-2010 dust-enshrouded state of the progenitor may not represent a quiescent stellar wind, but rather the remnant of a previous eruption episode. In this picture, the system undergoes repeated cycles of envelope disruption, a transient optically bright RSG-like phase, and subsequent re-enshrouding as newly formed dust obscures the donor. 

\subsection{An envelope interaction}
The products of binary interactions are predicted to show considerable diversity in their evolutionary outcomes, including extended supergiant phases and altered mass-loss histories \citep[e.g.][]{pauli2026drastic}. We speculate that the observed sequence of events in SN~2010da / ULX-1 can be explained by a brief interaction between the neutron star and the outer envelope of the donor star \citep[see e.g.][]{shishkin2020}. In this scenario, the neutron star temporarily grazed or penetrated the donor's envelope during a periastron passage or an episode of unstable mass transfer, triggering the 2010 outburst. The associated energy injection inflated the envelope, enhanced mass loss, and disrupted or destroyed the inner dusty environment. RSGs in this mass range are also susceptible to runaway radial pulsations in their late evolutionary stages, with surface velocities potentially exceeding the local escape velocity \citep{suzuki2025radial,sengupta2025dense}; such pulsation-driven instability may have compounded the effects of the binary interaction, further enhancing mass loss and envelope inflation during this phase. For ULX-1, this led to a prolonged phase of elevated accretion onto the neutron star, thus elevating the X-ray luminosity between 2010 and 2018, as well as causing the appearance of a cool, luminous donor.

However, the interaction did not result in a full inspiral or a common-envelope phase. Instead, the system subsequently detached, the donor envelope contracted, and dust formation resumed, restoring the system to an embedded, pre-outburst-like configuration with declining accretion from 2018 onwards. 

\subsection{Implications for the failed SN scenario}
\citet{chene2025ulx1} proposed that the donor star may have undergone a silent core collapse into a black hole in 2019, producing a ``failed SN.'' This scenario was motivated by the abrupt decline in optical brightness between 2018 -- 2024 and changes in the accretion regime. We note that models of failed SNe vary in their predictions regarding the expected amount of fading, timescale, presence of IR emission, and X-ray emission \citep[see discussion in][]{beasor2026ds1}, but in general a failed-SN outcome predicts: (1) a drastic drop in {\it bolometric} luminosity, (2) continued fading as accretion onto the nascent BH declines, and (3) no sustained IR output powered by a surviving stellar photosphere. ULX-1 stands in contrast to two other failed SN candidates, N6946-BH1 \citep{adams2017search, beasor2024BH1} and M31-2014-DS1 \citep{de2024ds1, beasor2026ds1}, which both appear to show a substantial drop in bolometric luminosity between the pre-disappearance and post-disappearance epochs. While the reason for this drop in luminosity is debated, given that ULX-1 has now returned to its pre-outburst luminosity and appearance, it would seem the system is in a different category the aforementioned failed SN candidates. For ULX-1, the failed SN scenario can be ruled out.

\subsection{Proposed timeline}
Taken together, the observations support an evolutionary scenario in which the donor alternated between a dust-enshrouded supergiant phase and a temporarily inflated, RSG-like configuration following episodes of strong binary interaction. We briefly outline a possible evolutionary sequence below: 
\begin{enumerate}
    \item \textbf{Pre-2010:} A dust-enshrouded supergiant with infrared-dominated emission.
    \item \textbf{2010 outburst:} Interaction between donor and NS clears inner dust and inflates the stellar envelope.
    \item \textbf{2010--2018:} Transient RSG-like phase with reduced obscuration and elevated accretion, culminating in the ULX-level luminosities that define this system. This is supported by the increasing X-ray luminosity through this time (see Fig.~\ref{fig:xraylc}).
    \item \textbf{2018--present:} Re-enshrouding as dust reforms and the system returns to an embedded state, accompanied by a marked decline in accretion rate. The \textit{Swift}/XRT monitoring demonstrates that ULX1 must have returned to at least sub-Eddington levels of accretion on average following this transition, even if a non-negligible level of accretion has persisted. The most recent firm X-ray detection --- a \textit{Chandra} observation in 2020 (MJD 58965) yielding $L_\mathrm{X} \sim 10^{37}$\,erg\,s$^{-1}$ \citep{chene2025ulx1} --- is consistent with all subsequent monthly XRT upper limits, meaning we cannot yet constrain whether the source has settled into a persistent low-luminosity state at this level or has faded further toward quiescence. Either way, both scenarios represent a substantial drop from the peak accretion rate responsible for the ULX-phase luminosities (see Fig.~\ref{fig:xraylc}).
\end{enumerate}

\section{Conclusions}\label{sec:conclusions}
We have presented new $JWST$ NIRSpec and MIRI IFU spectroscopy of
SN~2010da / NGC~300 ULX--1, revealing that the system remains
luminous but is now heavily obscured by warm, optically thick dust.
The disappearance of molecular absorption features associated with
the previously observed RSG-like spectrum, combined with only a modest
decline in bolometric luminosity, strongly indicates that the donor star
has survived and returned to its pre-2010 outburst state.

Our \emph{JWST} observations rule out a failed supernova scenario, which would predict a far more dramatic and sustained decline in bolometric luminosity. Instead, the data point to a major, non-terminal event that temporarily triggered by interaction between the NS and the donor star, that altered the circumstellar environment and accretion state of the system. The rapid photometric and spectroscopic evolution observed over the past 16 years underscores the importance of continued multiwavelength monitoring. Given that this is a binary system, we may expect further outbursts analogous to the 2010 event in the future, which would provide conclusive confirmation of this scenario. SN~2010da/ULX-1 may be analogous to the quasi-periodic SN impostors SN~2000ch \citep{aghakhanloo2023200ch} and AT~2016blu/NGC~4559ot \citep{aghakanloo2023AT2016blu}, which exhibit recurrent eruptive behaviour driven by episodic mass loss and circumstellar interaction, albeit on shorter timescales likely set by their more compact progenitor systems.

\section*{Acknowledgements}
We would like to thank the anonymous referee whose comments helped improve the paper. The authors would like to thank Sebastian Kamann for useful discussions on working with IFU data and Kishalay De for providing NEOWISE lightcurves at the proposal stage. The authors would also like to thank André-Nicolas Chené for useful discussions. E.R.B. is supported by a Royal Society Dorothy Hodgkin Fellowship (grant no. DHF-R1-241114). D.J.W. acknowledges support from the Science and Technology Facilities Council (STFC; grant code ST/Y001060/1). This work is based on observations made with the NASA/ESA/CSA James Webb Space Telescope. The data were obtained from the Mikulski Archive for Space Telescopes at the Space Telescope Science Institute, which is operated by the Association of Universities for Research in Astronomy, Inc., under NASA contract NAS 5-03127 for JWST. These observations are associated with program 3738.

\section*{Data Availability}
The $VLT$ X-Shooter spectra are available via the ESO Data Archive. The $JWST$ data are available via MAST (doi:10.17909/p3m4-s245).



\bibliographystyle{mnras}
\bibliography{references} 





\bsp	
\label{lastpage}
\end{document}